# A RESTORATION IN THE TIME SYMMETRIC THEORY AND ASSOCIATED NEW RESULTS

**Evangelos Chaliasos**

365 Thebes Street

GR-12241 Aegaleo

Athens, GREECE

*Abstract*

The concept of "absorption" of a photon moving backwards in time is reexamined, and it is found that its interpretation as absorption is wrong, its original interpretation rather as an emission being restored. The result is that the anticosmos is invisible rather than indistinguishable from the cosmos, and as such it is qualitatevely a candidate for constituting the dark matter. Newton´s law of gravitation is then generalized to refer either to matter (of the cosmos) and/or antimatter (of the anticosmos), and quantitatevely is found for the anticosmos to constitute the dark matter, as being (absolutely) equal in ammount of antimatter to the ammount of matter in the cosmos.Finally, a mechanism of gravitational repulsion from antimatter is proposed to explain the observed deceleration of *Pioneer 10* and *Pioneer 11,* and an analytical way is exposed to calculate the position, as well as its (anti-)mass, of the responsible anti-body.

## 1. Introduction

The author developed a complete theory of time symmetry during the years 1997-1999. The first draft of the theory, divided into 12 papers, was written in 1998-1999, but it was not worked out, and moreover it remained unpublished, up to now. Nevertheless some applications of the theory can be found in [1, 2, 3, 4]. In these references the original ideas, mainly concerning the emission and absorption primarely of photons and secondarily of particles and antiparticles, were modified. I see now that those modifications were wrong. Thus I came back to the original form of these ideas, as it can be seen in section 2 of [5], as well as in an appendix of a paper submitted for publication, where I summarized the basic points of the theory. Similarly, that summary appeared in Greek [6], but unfortunately it was based on the (wrong) modified ideas. Thus, I considered it necessary to reject these (wrong) modified ideas, making a restoration of the whole theory and thus coming back to the original ideas, by writting the present theory. In this paper also the very important implications of that restoration are included. The reader can thus appreciate by himself their significance.

The "absorption" of a photon, of energy $-h\nu$ and moving backwards in time, interpreted as a real absorption is rejected, and, instead, its interpretation as an equivalent *emission* of the photon, with energy $h\nu$ and moving forward in time, is proposed in section 2. In the same section the invisibility of the anticosmos is proposed, and the construction of an emission telescope, able to see the otherwise invisible anticosmos, is suggested. It is also proposed to identify that invisible anticosmos with dark matter, in the same section, and it is pointed out that the problem of dark matter can be solved qualitatevely.

In section 3 the non-equivalence of an antiparticle of mass –m and moving backwards in time with a normal particle of mass m and moving forward in time is proposed, and thus my first cosmological model (unpublished), based on an equivalence, is rejected. Newton´s law of universal attraction is then generalized to a law of attraction as well as repulsion between mass and/or anti-mass. Based on this generalized Newton´s law a mechanism is proposed by which the problem of dark matter is solved quantitatevely as well, in the same section.

Finally, in the 4$^{th}$ section, the puzzle of *Pioneer 10* and *Pioneer 11* experiencing an additional (besides Sun´s) deceleration is solved by assuming the existence of an invisible anti-body beyond, but close to, the Solar system, which repels either spacecraft. An analytical method is then proposed of how to determine the position and the anti-mass of the anti-body in the same section.

2. **Some aspects of emission & absorption of photons**

Suppose that a photon of energy $-h\nu$ and moving backwards in time is incident on a human eye, or on a telescope. Suppose that it is "absorbed" in a time interval $-t$, evidently negative. The incident flux is then $\Phi = (-h\nu)/(-t)$. If we perform the division using the usual algebra, we will find $\Phi = h\nu/t > 0$, and thus we would conclude that the photon would be absorbed, as in the case of photons with positive energy and moving forward in time. But this is *incorrect*. The reason is that we have to use *antialgebra* rather than usual algebra in the division, since the photon is a normal photon in the anticosmos. Thus, the correct result for the flux is $\Phi = -h\nu/t < 0$, and the photon will actually be *emitted*, as long as the human eye, or the telescope, can *emit*. But this is not the case, so we will see *nothing*. We would see the photon only if we were belonging to the anticosmos, in which case we would have $\Phi = (+h\nu)/(+t) > 0$ and the photon would then be really absorbed! Of course opposite things will hold concerning "emission" of photons of negative energy –hv and moving backwards in time.

Thus, finally we can say that photons with negative energy and moving backwards in time are *equivalent* to photons with positive energy and moving forward in time. This is in complete accordance with the fact that photons coincide with their own antiparticles.

As a result of the fact that photons of negative energy and moving backwards in time cannot be traced apparently by human beings, and since such photons are the ones coming from the anticosmos, we can conclude that *the anticosmos is invisible.*

The above conclusion is true as long as telescopes cannot emit. We are thus oblidged to *invent emission telescopes,* that is telescopes whose emulsion used in them can emit. In the atoms of usual emulsions the electrons are in their ground states, being thus unable to emit. It is evident then that we have to invent "emulsions" in the atoms of which the electrons are in excited states, and able to emit in the frequency of the incident photons jumping to their ground states. This could be achieved possibly by using fluorescencing materials, or laser-like devices since we know that in lasers we can achieve the desirable population inversions.

Finally, our minds go directly to dark matter, when we are thinking of an invisible anticosmos. This qualitative capture has of course to explain also the quantitative feature of dark matter of being equal in ammount to the visible Universe, in order to be valid. We will see that this is indeed the case, in the next section.

### 3. Generalized Newton´s law and dark matter

Since a particle is well distinguished from its antiparticle, it is evident that an antiparticle of mass –m and moving backwards in time *is not* equivalent to the "same" particle presumedly with mass m and moving forward in time (as in the case of photons).

The first cosmological model proposed by the author (unpublished) was based on the assumption of such a hypothetical equivalence. We then chose the closed Friedmann model among the totally three Friedmann models (closed, open, and flat), because only this model is symmetric under a time reversal. This now can be seen to be in error. That is why I was oblidged (besides the small age of the Universe resulted) to propose a second cosmological model [5].

If we accept the existence of bodies (anti-bodies) of negative mass (and moving backwards in time), the question arises of how the Newton law of gravitation has to be modified in the case of such bodies (anti-bodies) of negative mass being involved. We propose that law remains *unchanged,* the only difference being that now we have to admit *negative* masses to be introduced in the usual formula expressing the law, with the result that we could take gravitational *repulsion* besides attraction. Of course we have to consider the resulting *accelerations* rather than the forces themselves, since these two physical quantities may have opposite directions. We thus find finally, using either the usual algebra or antialgebra, that *matter attracts* (either matter or anti-matter), while *antimatter repels* (either matter or antimatter again).

Suppose now that we have (absolutely) equal ammounts of matter and antimatter arranged uniformly on a (straight) line (we consider one dimension merely for simplicity). We can achieve this by simply considering a series of bodies and anti-bodies of (absolutely) the same mass placed successively in equal distances from one another on the line (see fig. 1). Suppose also that a test *particle* of mass μ is placed in the middle O of the distance of two successive of them M and A (see fig. 1). Then M *attracts* μ with an acceleration $a_m$ , while A *repels* μ with an acceleration $a_{-m}$ . Because we have assumed that OM = OA, it can be seen that also $a_m = a_{-m}$. Then, if A (-m) is invisible, we will attribute to M (m) an *attraction* $a = 2a_m$ acting on O (μ). Consequently we will attribute to the visible M, plus an additional but unknown (invisible!) ammount of *matter* (the dark matter!) a mass *2m rather than m.*

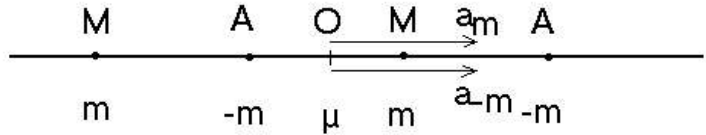

Fig. 1

It is thus obvious that if we *identify the (invisible) anticosmos with dark matter,* we will attribute to the cosmos (thinking of it as being the whole Universe) a doubled ammount of matter. In other words the mass of the Universe derived by dynamical methods, will be two times [8] the mass of the visible Universe, in this way the puzzle of dark matter being resolved even *quantitatevely.*

4. **Gravitational repulsion acting on *Pioneers***

A very strange effect was observed lately. Namely an additional force (besides the Sun´s one) was observed which decelerated the spacecrafts *Pioneer 10* and *Pioneer 11* at the limits of the Solar System. It was like an invisible object beyond, but close to, the Solar System repeled them [7].

However, it is easy to explain the observations if we assume that an *anti-body* (made up of anti-matter) was responsible for this repulsion! In fact, we can even determine its position and its (anti-)mass from the observations, by using simple mathematics, as follows.

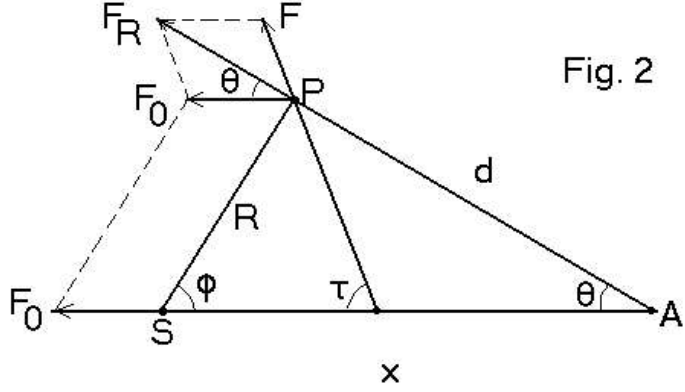

Fig. 2

Suppose that we are on the Sun at S, the *Pioneer* is at P, and the anti-body, of absolute mass m, is at A (fig. 2). Suppose that A repels S by $F_0$, and P by $F_R$. The distances are SA = x, SP = R, and AP = d, while the angles are as in the figure. We will proceed as in the case of the tides. If we supposse that we have subtracted the Sun´s attraction acting on P, then we have to also subtract from the repulsion $F_R$ the repulsion $F_0$ acting on the Sun. The result will be the force F acting on P, the Sun´s attraction acting on P not arrearing in our analysis. Thus, first of all we assume that we observe the force F acting on P, so that A lies on the plane (F, S). The distance R is also known.

From the figure we have immediately the four trigonometric relations

$$\frac{x}{\sin(\varphi+\vartheta)} = \frac{d}{\sin\varphi} = \frac{R}{\sin\vartheta}, \tag{1}$$

$$d^2 = R^2 + x^2 - 2Rx\cos\varphi, \tag{2}$$

$$R^2 = d^2 + x^2 - 2dx\cos\vartheta. \tag{3}$$

Then the two relations

$$\tan\vartheta = \frac{F\sin\tau}{F_0 + F\cos\tau}, \tag{4}$$

$$\tan(\tau - \vartheta) = \frac{F_0 \sin\tau}{F + F_0 \cos\tau}, \tag{5}$$

provide another system of two simultaneous equations in the two unknowns $F_0$ and $\tau$. Thus finally, if we solve it, we will be able to use the equation

$$F_0 = Gm/x^2, \tag{6}$$

from which we can find m.

In this way we can determine the position of A, since we will have found φ and x, and its absolute mass m.

**Appendix A**

We have seen that a photon of negative energy –hν moving backwards in time is equivalent to a photon of positive energy hν moving forward in time, concerning absorption and emission. Nevertheless, since a photon coincides with its own antiparticle, this picture is not completely satisfactory. It would be more satisfactory if its energy was zero. But, since during the absorption and/or the emission we observe energy difference hν, and by virtue of the above equivalence, we are led to propose the picture of the photon consisting of two parts: one of energy hν/2 moving forward in time, and one of energy –hν/2 moving backwards in time, both following the same world line but in opposite directions, in such a way that they occupy the same position at the same time. In this case, on the one hand the total photon´s energy would be zero as required, and on the other hand the photon would be absorbed and/or emitted by the energy ammount hν/2 + hν/2 = hν!

Note that we cannot observe the photon´s energy otherwise, but only during its absorption and/or emission. Thus, the Planck´s picture of light quanta remains valid.

On the contrary, Einstein´s assumption of these quanta *travelling* in space as photons has to be modified as proposed above.

**Appendix B**

A simple way to obtain at the same time both a matter (cosmos) and antimatter (anticosmos) photograph is just to place in the telescope an illuminated fluorescencing plate before (and attached) to a usual photographic plate! Why? First the illumination will maintain the fluorescencing material in a state of partial excitation. Then the positive photons (moving forward in time), coming from the visible (matter-)Universe (cosmos), will excite more atoms of the fluorescencing material where they are incident, while the negative photons (moving backwards in time), coming from the invisible (antimatter-)Universe (anticosmos), that is from dark matter, will reduce the number of the atoms of the fluorescencing material being in excitation where they are incident, since they will lead these atoms to their ground states (because right of their negative energies!). The result will be bright points (stars!) and dark points (anti-stars!), in a gray background, to appear on the fluorescencing plate, which then will be imprinted on the photographic plate!

It is evident that in this way we can also observe directly the sky (visible part plus dark matter) on the fluorescencing plate (screen!), if the photographic plate is absent!

Needless to say that in this way we can observe either the cosmos or the anticosmos alone by just decreasing or increasing respectively the illumination, adjusting in this way the gray background to be white or black respectively.